# Software Design Principles of a DFS Tower A-CWP Prototype

Felix Schmitt, Ralf Heidger, Stephen Straub, and Benjamin Weiß

DFS Deutsche Flugsicherung GmbH
SH/P Systemhaus, Am DFS - Campus 7, 63225 Langen, Germany
email: felix.schmitt@dfs.de

*Abstract* – **SESAR is supposed to boost the development of new operational procedures together with the supporting systems in order to modernize the pan-European air traffic management (ATM). One consequence of this development is that more and more information is presented to –and has to be processed by– air traffic control officers (ATCOs). Thus, there is a strong need for a software design concept that fosters the development of an advanced (tower) controller working position (A-CWP) that comprehensively integrates the still counting amount of information while reducing the data management workload of ATCOs.**

**We report on our first hands-on experiences obtained during the development of an A-CWP prototype that was used in two SESAR validation sessions.**

## I. INTRODUCTION
## SESAR AND "ATS COMPONENTWARE"

In the recent decade the amount of air traffic has reached a point where a simple increase of manpower is no longer sufficient to guarantee safe and efficient air traffic services (ATS). Novel operational procedures together with new or improved supporting systems need to be developed. Furthermore, information sharing between all stakeholders must be extended and simplified in order to foster collaborate decision making (CDM).

Often, new systems that support a specific operational workflow or that simply present information to ATCOs are introduced via an additional display and additional input devices. Consequently "modern" towers tend to become cluttered workplaces where information has to be searched and entered at various places distracting ATCOs to focus on their genuine job, namely air traffic control. One reason for this is that today's ATS systems are mostly monolithic and inflexible solutions that lack of easily accessible interfaces. More important, they lack of a system design that allows for convenient adaptations and extensions in the first place. European air navigation service providers (ANSPs) are aware of that, and one platform for their joint effort to change the status-quo is the Single European Sky ATM Research Programme (SESAR). A central aim of European ANSPs that participate in SESAR (the A6 group) is the evolution of ATS systems towards increased flexibility by defining standardized interfaces between standardized functional components. In the A6 group this idea is promoted under the keyword "ATS Componentware".

It is the policy of the A6 group that four high-level requirements shall be considered in future ATS systems [1]:
- separation of information provision/consumption
- loose system coupling
- using open standards
- using service oriented architecture.

## II. THE BIG PICTURE

Being only a subset of a complete ATS system, it is helpful to see how a CWP is embedded into a bigger scope. A decomposition of a component based ATS system is shown in Figure 1. Following a V-model decomposition [2], the whole ATS complex is considered being a system of systems. A system incorporates the infrastructure of a specific ATS domain, e.g. a tower or an area control centre. These systems are further divided into sub-systems, e.g. a primary system and a fallback system each of which having segments like sensor data processing system (SDPS), flight data processing system (FDPS), etc. Each segment comprises hardware, software, and network. Software is built up of components, i.e., executable processes.

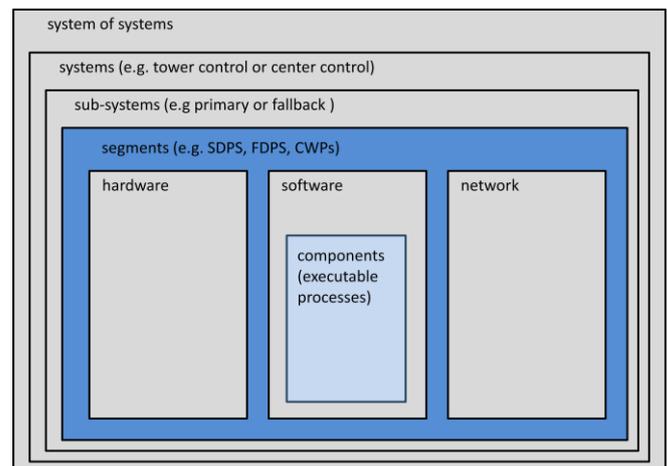

Figure 1 Decomposition of a component based ATS system. The description of our prototype starts at the segments layer (dark blue) and focuses on the communication between components (light blue).

Following the above presented decomposition, a generic and component based ATS sub-system can be depicted as shown in Figure 2. Different segments are vertically aligned and each of which is horizontally divided into an interface level, a server level, and a client level. The interface level encapsulates communication with external systems that offer propriety interfaces only. The generic segment shows how the components A and B (see Figure 1) are embedded in this view.

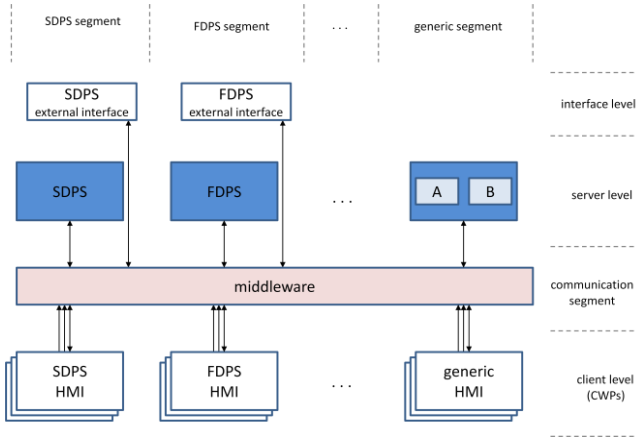

Figure 2 Architecture of a generic ATS sub-system. Dashed vertical lines separate different segments. Note that the listing of segments is non-exhaustive. The generic segment is further decomposed to software components A and B. Dashed horizontal lines show the different levels and the communication segment.

It is obvious that the communication segment, i.e., the middleware plays a central role in this architecture.

In the following we introduce the initial scope of the DFS A-CWP prototype starting our description at the segments layer shown in Figure 1. We will present the software design principles that we employed in order to realize a prototype that integrates different segments from different manufacturers, each of which are separately in operational use.

### III. SCOPE

DFS is a member of SESAR project "Integrated Tower Working Position" [3] aiming at the definition and development of a component-based A-CWP with an integrated data management. The ultimate goal would be a seamlessly integrated A-CWP that is made up of components –possibly from different manufacturers– yet completely hiding its fractional interior from the user.

Development in this context does not mean a development from scratch, but an integration of already existing products from different manufacturers. The baseline for the DFS A-CWP prototype was:

- PHOENIX/SDPS, a combined air/ground situation display and tracker [4] together with several safety net components [5] developed by DFS
- SHOWTIME/FDPS, an electronic flight strip system [6] developed by DFS
- smartStrips/FDPS, an electronic flight strip system developed by Frequentis

The combination of PHOENIX and smartStrips was used in a validation in June 2012 at DFS tower simulator in Langen (Germany) [7] and the combination of PHOENIX and SHOWTIME was used in a validation in November 2012 at tower Hamburg (Germany) [8].

Following the nomenclature introduced in section 2, each of the three products is generically referred to as a *segment* of the overall prototype. Regardless of the segment, a unit with a well defined input, output, and functionality is called a *component*. The term *message* is used to describe an atomic piece of information, e.g., a data packet that is transmitted between components.

### IV. DESIGN CHOICES

A closer look at architecture and platform of the three different segments already reveals some challenges for the integration. PHOENIX is a multi-process system that uses a propriety middleware and a SQLite database. It is developed in C++ under a Linux platform. SHOWTIME is a client-server system based on an Oracle SQL database. It is developed in C++ and Python under a Linux platform. smartStrips is a client-server system based on a PostgreSQL database. It is developed in Java under a Microsoft Windows platform.

*Message Oriented Middleware*

The first and most fundamental decision is to use a Message Oriented Middleware (MOM) as the core of the integration. One possible alternative would have been a database-centric architecture, however, the concept of a MOM allows for a natural evolution of a component-based architecture – sometimes called a service-oriented architecture [9] – and is in line with the architecture shown in Figure 2. Since the prototype is jointly developed by different manufacturers, there are constraints on the choice of a message broker, i.e., an actual instance of a MOM. In order to prevent mutual dependencies between the different the manufacturers, we decided to employ Apache's ActiveMQ [10] which is an open-source MOM that offers APIs for C++, Python, and Java amongst others. Using ready-made APIs also spares us from spending time for the implementation of low-level transport layers.

*Publish/Subscribe*

ActiveMQ offers the publish/subscribe pattern and from our experience with the internal middleware of PHOENIX we know that a publish/subscribe pattern is a convenient way to foster the design of a loosely coupled architecture, because developers are encouraged to think asynchronously and to use a non-blocking design from the very beginning of the development process. Of course, some components, usually parts of the HMI, will always require request/reply patterns.

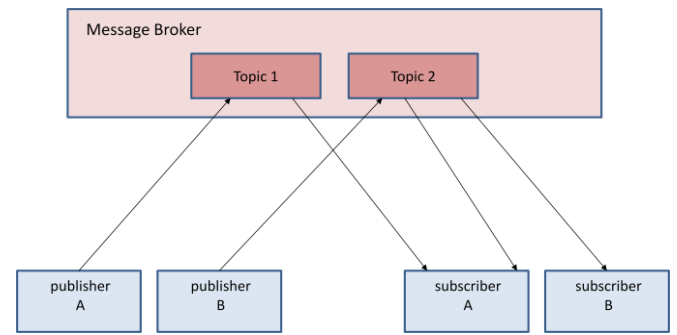

Figure 3 Topics gather information of the same type like flight plans, QNH values, etc. There is no need to distinguish publisher and subscriber components from different segments.

In order to clearly arrange the message flow between components, we make use of topics [10]. As shown in Figure 3, a topic gathers messages of the same type of information, e.g., one topic for "QNH", one topic for "Flight Plans", etc.

*Data Consistency*

Whenever data is manipulated by different components, data-consistency through all segments must be assured. To this end, we define a dedicated data owner for each message type, e.g. the data owner for flight plans could be a flight-plan database. A contribution/publication pattern as depicted in Figure 4 is used to communicate changes of the data.

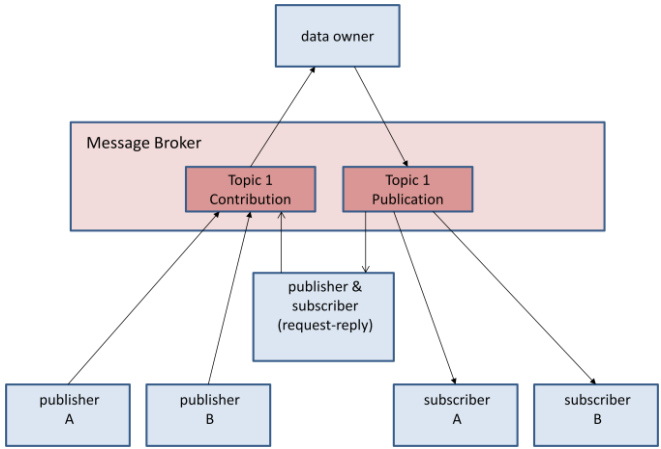

Figure 4 In order to assure data consistency, the data owner has an exclusive subscription-right to "Contribution" and an exclusive publication-right to "Publication".

Components that need to manipulate data that they do not own publish their information on the corresponding contribution topic. Only the data owner is allowed to subscribe to that topic. After internal processing of the data, the data owner has the exclusive right to publish the new information on the corresponding publication topic. A timely sequence of an exemplary message flow is shown in Figure 5.

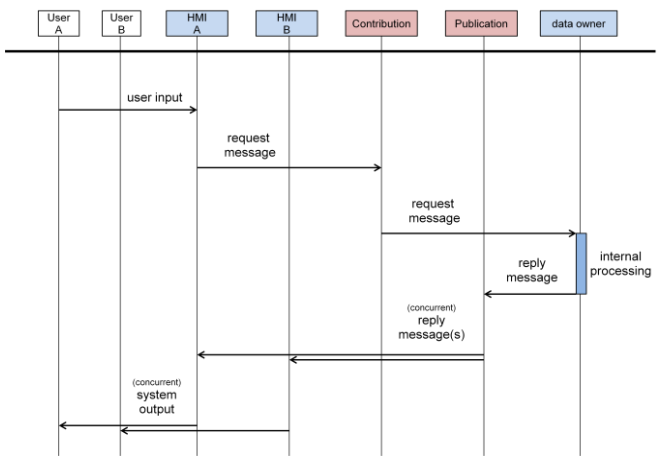

Figure 5 A timely sequence of a message flow is shown. For example, the user input could be the change of a flight plan which first has to be processed by the flight plan database and then has to be published to all CWPs. Small vertical distances between the arrows shall denote concurrent message flows.

*Plain-text Messages*

Messages that are transmitted between ATS components are often in a binary format like Eurocontrol's Asterix [11] because when they were originally defined, there were strong limits on bandwidth and processing power and performance prevailed human readability. However, using messages that are not human readable slows down debugging especially in more complex environments where different kinds of messages are used. Since modern IT has overcome these performance constraints, the central requirement on the message format should be convenience for the developers. Debugging and black-box testing is extremely simplified when there is no need to start with the development of message translators or test-message generators. Therefore, our prototype employs solely plain-text, human readable messages for information exchange.

*Message Validation*

Syntactical validation of messages at each input and each output channel strongly improves maintainability. XML is a powerful human-readable and machine-readable plain-text message format that comes with a comprehensive tool chain for most platforms and languages. In particular, XML offers syntactical validation of messages against XML Schema Definitions (XSDs). Once agreed on XSDs for all messages types, inter-communication problems that are often not detected before the final integration of all components from all segments can be prevented.

*Hierarchy of Brokers*

In a multi CWP environment only the client (HMI) components are really "multi". Neglecting redundancy, the server components are typically unique. Our design reflects the separation between clients and servers already at broker level. As depicted in Figure 6, we use a central message broker for the server components and a local message broker for each CWP client.

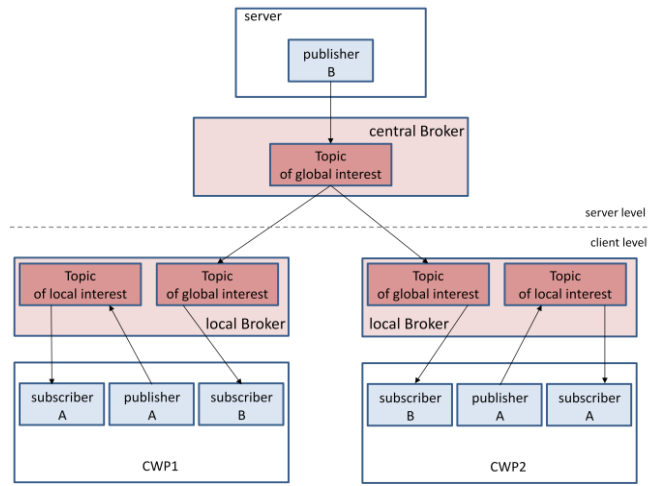

Figure 6 The DFS A-CWP prototype employs a hierarchy of message brokers. Note that it is irrelevant to which segments the publishing and subscribing components belong.

Using not a single message broker but a hierarchy of message brokers introduces lots of flexibility. One can use the same topic names even for information that is specific for each CWP client, e.g., the selection of a target or the acknowledgement of a safety alert. Furthermore, attaching an additional CWP client, or any other component, is easy since no server component needs to know the number and location of all client components. (For message routing we employ Apache Camel, an open-source message routing and transformation engine [12].)

*Error Recovery*

In order to assure message delivery, a publishing component usually blocks until acknowledgements from all receivers have arrived. This would inherently introduce coupling between components. Therefore, we define a global strategy to handle failed message-delivery. Under the assumption that a local broker is as reliable as any other local component, the responsibility of message delivery can be delegated to the local broker. As depicted in Figure 7 dead letter topics are used to release publishing components form the duty of observing a correct message delivery. A dead letter topic gathers messages that were not retrieved by all subscribers in a predefined time period. A dedicated component is then responsible for error recovery. This mechanism also allows for a basic debugging of components from foreign segments. Of course, this mechanism can be used for other sorts of error recovery as well. This is a convenient way to make a component more asynchronous.

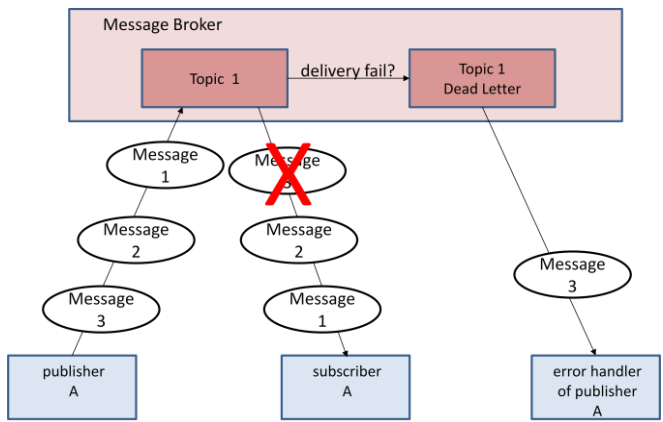

Figure 7 Local brokers can be used to release publishers from the duty of observing correct message delivery. In the above case, message 3 is not retrieved by subscriber A in due time and therefore forwarded to a dead letter topic.

## V. CONCLUSION

Using a Message Oriented Middleware, as the core of the integration of the DFS A-CWP prototype turned out to be a good choice since the high-level requirements presented in the introduction are almost naturally fulfilled in the resulting architecture. Through a message oriented middleware, a high degree of flexibility on all levels shown in Figure 2 is assured. One can easily add new server and HMI components without any side-effects on the existing functionality. External systems that offer only propriety interfaces can be attached by using dedicated interface agents, i.e., components that act as translators. The use of publish/subscribe and contribution/publication patterns leads to a loose coupling between components with a clear separation of information provision and information consumption. Since ActiveMQ implements the Java message service (JMS) standard and we intend to publish the XSDs that were developed for this prototype, a first step towards the definition open standards for the A-CWP has been done.

It should be noted that the gain in flexibility that comes with an integration based on a message oriented middleware is not free of disadvantages. In a database-centric integration data is represented twice, once in the database and once in the component that uses that data. A middleware-centric integration introduces a third data representation at the communication layer. Furthermore, the introduction of a new component, i.e. the message broker, to the existing software architecture also introduces new risks in terms of stability. If one can do without the rich functionality of ActiveMQ, it could be worthwhile to have a look at a broker-less solution like ZeroMQ [13] or Crossroads I/O [14].

## VI. ACKNOWLEDGEMENT

This work was co-financed by the European Union and EUROCONTROL in the framework of SESAR. We would like to thank Wolfgang Schneider and Radoslav Natchev for their advices and many fruitful discussions. Many thanks to Apache Software Foundation for offering great solutions like ActiveMQ.